# Deep Learning in Infant Functional Neuroimaging: Challenges, Advances, and Future Directions

Dan Hu[1,2†], Jiale Cheng[1,2,3], Weiran Xia[1,2,3], Kangfu Han[1,2], Matthew Wu[2], Li Wang[1,2], Weili Lin[1,2], Gang Li[1,2†],

1. Department of Radiology, University of North Carolina at Chapel Hill, Chapel Hill, NC, USA
2. Biomedical Research Imaging Center, University of North Carolina at Chapel Hill, Chapel Hill, NC, USA
3. Lampe Joint Department of Biomedical Engineering, University of North Carolina at Chapel Hill and North Carolina State University, Chapel Hill, NC, USA

† Corresponding authors:

Gang Li, gang_li@med.unc.edu

Dan Hu, danhu@email.unc.edu

**Highlights**

- Deep learning shifts infant fMRI from group mapping to individualized prediction.
- FMRI representations strongly shape biological interpretation and model validity.
- Deep models can capture functional organization, trajectories, and clinical risk.
- Robust validation must address generalization and biological convergence.
- Future progress needs multimodal, interpretable, developmentally grounded models.

**Abstract**

Infancy is a critical developmental window characterized by rapid functional brain reorganization, during which large-scale networks emerge, individualized connectome signatures continue to form, and early deviations may shape long-term cognitive and clinical outcomes. Functional MRI (fMRI) offers an opportunity to study these processes in vivo, yet extracting developmentally meaningful information from it remains challenging due to comparably short scan duration, structured motion artifacts, variable scan states, and rapid brain maturation. Amid these challenges, deep learning has expanded the capacity of computational neuroimaging by learning robust representations from noisy, high-dimensional data, integrating complex spatial and temporal information, and capturing the nonlinear and rapidly evolving organization of the developing brain. Here, we review recent advances in deep learning for infant functional neuroimaging, synthesizing progress across input representation formatting, population and individualized brain mapping, longitudinal trajectory forecasting, robust and explainable model evaluation, and biological translation. Collectively, these methodological advances mark a paradigm shift in infant functional neuroimaging from descriptive, group-level analyses toward reliable, individualized, and developmentally grounded models. Future progress will depend on larger and more diverse longitudinal datasets, developmentally appropriate model designs, rigorous and standardized evaluation, and stronger integration of computational predictions with biological mechanisms towards clinically meaningful outcomes. Addressing these priorities will help establish deep learning as a robust framework for understanding early functional brain development, identifying meaningful developmental variation at the individual level, and ultimately supporting earlier and more precise assessment of neurodevelopmental risk.

## 1. Introduction

The first two years of human life represent a remarkably formative period for functional brain organization, marked by rapid changes in network topology, integration, and differentiation (Gao et al., 2013, Gao et al., 2015, Zhang et al., 2019). Large-scale functional architecture is already discernible around birth and undergoes substantial reorganization throughout infancy (Gao et al., 2009, Doria et al., 2010). Functional MRI (fMRI) offers an indispensable, non-invasive means of tracking this reorganization in vivo. Compared with electroencephalography (EEG) and functional near-infrared spectroscopy (fNIRS), which provide high temporal resolution or portable assessment but have limited spatial coverage and access to deep brain structures, fMRI provides whole-brain coverage of distributed functional networks, making it especially suited for mapping infant connectome organization and relating early functional variation to cognitive, behavioral, and clinical outcomes (Smyser et al., 2010, Cao et al., 2017, Agyeman et al., 2023, Hu et al., 2022). Infant functional neuroimaging therefore provides a unique opportunity to identify early neural variation with potential relevance to developmental prognosis and timely intervention.

Conventional analytical approaches have established much of what is known about infant functional brain organization. Seed-based connectivity and independent component analysis (ICA) revealed the early emergence of coherent functional systems; graph-theoretic measures characterized changes in integration, segregation, and topology; infant-specific parcellations, developmental growth charts, and functional gradients refined descriptions of normative spatial organization and brain maturation (Cao et al., 2017, Wen et al., 2019, Wang et al., 2023a, Yuan et al., 2024, Yin et al., 2025). These approaches remain indispensable by providing biologically interpretable references and defining properties that advanced computational models should recover, preserve, or explain. However, their analytical constraints increasingly limit the questions they can resolve. Many conventional analyses rely on group averages, predefined seeds or atlases, static connectivity summaries, hand-crafted features, and relatively simple models of age-related changes. Although effective for identifying broad developmental patterns, these strategies can obscure individual organization, compress temporally rich signals into fixed summaries, and inadequately represent nonlinear and heterogeneous trajectories.

The field is therefore moving beyond describing average network development toward determining whether an individual infant's functional organization can be mapped reliably, separated from fixed measurement prior, and linked to brain and behavioral outcomes. Meeting these goals is particularly challenging because infant fMRI is typically short, noisy, sparsely sampled, and heterogeneous across age, scan state, protocol, and site. Atlas-based parcellation can delineate broad functional territories, but rapidly changing cortical geometry and topography may limit infant- or subject-specific mapping (Wang et al., 2023a); graph measures can summarize population-level network maturation (Wen et al., 2019, Jiang et al., 2023) but do not readily resolve individualized developmental trajectories; conventional statistical and machine learning model can predict outcomes from selected features (Smyser et al., 2016, Ball et al., 2016, Li et al., 2025, Yu et al., 2022a) but are often constrained by fixed feature representations and relatively inflexible model structures, limiting their ability to capture complex nonlinear interactions among distributed functional features. More broadly, these methods are not naturally designed to learn multilevel representations from high-dimensional, longitudinal, and multimodal observations. Addressing these questions requires extending conventional analysis with models capable of representing the complexity, individuality, and rapid dynamics of early brain development.

Deep learning offers such a framework by learning hierarchical representations directly from images, time series, and connectomes rather than relying exclusively on predefined features or fixed summaries. Its nonlinear structure can integrate spatial and temporal information, accommodate age-dependent relationships, and model developmental, individual, and measurement-related variation. Autoencoders and variational models can compress functional connectivity and activity patterns into latent representations of developmental and individual differences (Hu et al., 2020b, Kim et al., 2023b, Kim et al., 2024). Graph neural networks preserve connectome topology rather than flattening connectivity into unordered vectors (Li et al., 2022, Zhang et al., 2022), whereas Transformers and attention models capture context-dependent interactions across regions, time windows, pathways, and modalities (Fang et al., 2023, Mao et al., 2024). Self-supervised, contrastive, and generative approaches can exploit unlabeled data, improve representations from limited observations, harmonize heterogeneous acquisitions, and support individualized forecasting (Tu et al., 2025, Xia et al., 2026b, Cheng et al., 2025b). Deep learning may therefore advance not only prediction, but also individualized functional mapping, developmental trajectory modeling, and the identification of early neural variation linked to later outcomes.

In this Review, we synthesize recent advances in deep learning for infant functional neuroimaging across interrelated domains as demonstrated in **Figure 1**: (1) infant fMRI data and representations; (2) functional brain mapping; (3) modeling developmental dynamics and individual trajectories; (4) translational applications, model evaluation, interpretability, and robust learning. We conclude by examining open challenges and future priorities, arguing that deep learning is most valuable when it advances developmental cognitive neuroscience rather than benchmark performance alone. Ultimately, deep learning can serve as a powerful integrative framework that builds on conventional analytical approaches for understanding early functional brain development, identifying meaningful developmental variation at the individual level, and supporting earlier and more precise assessment of neurodevelopmental risk.

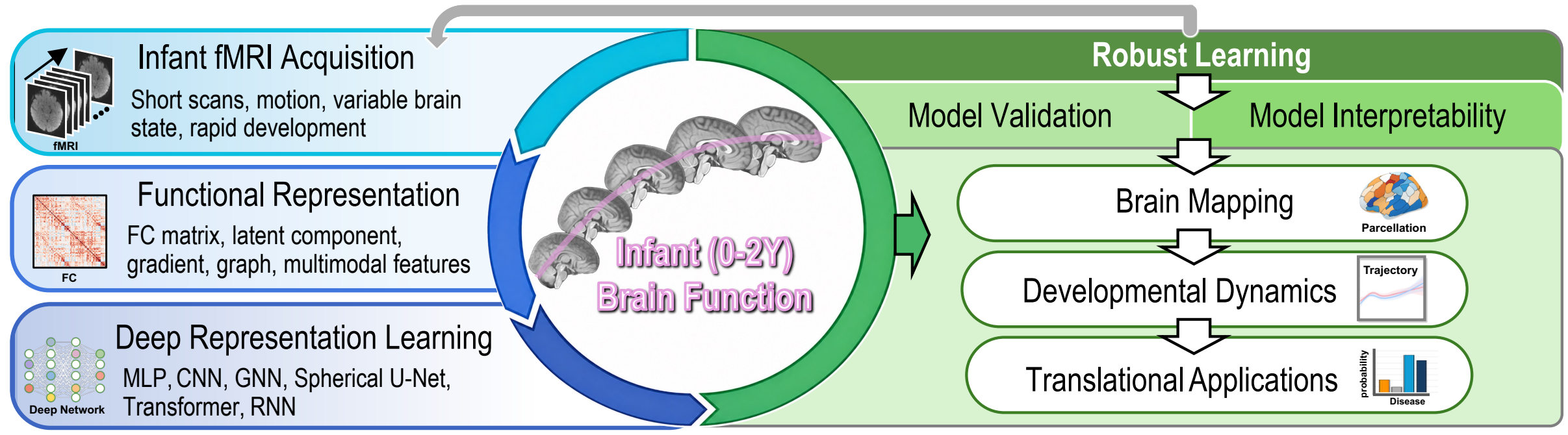

**Figure 1. Conceptual framework for deep learning in infant functional neuroimaging.** Infant fMRI acquisition produces developmentally constrained functional data that are transformed into functional representations and embedded through deep representation learning to characterize baby brain function, supporting brain mapping, developmental-dynamics modeling, and translational applications. Robust learning addresses the specific challenges of infant functional neuroimaging and serves as an overarching requirement for strengthening model validation and interpretability across downstream analyses.

## 2. Review Scope and Literature-Search Strategy

This Review characterizes the conceptual and methodological landscape of deep learning in infant brain functional MRI research. The primary population of interest was human infants from birth to two years of age (0-2Y). Fetal studies were included when providing essential prenatal or developmental context, whereas studies of older children were considered selectively if they reported infant-specific findings or introduced methods directly transferable to infant functional neuroimaging.

Resting-state fMRI was the principal modality of interest, supplemented by task-based fMRI and selected multimodal studies when they directly informed functional mapping, prediction, validation, harmonization, or biological interpretation. Methodologically, we focused on deep learning and closely related predictive or developmental modeling frameworks. Conventional machine-learning studies were cited selectively to establish historical context, provide benchmark comparisons, and clarify the additive value of deep learning.

Systematic searches were conducted in PubMed/MEDLINE, Scopus, Embase, and IEEE Xplore on 21-22 July 2026 for records published from 2010 onward, without language restrictions. Search strategies combined terms related to early development, fMRI and fMRI-derived functional measures, and computational modeling. Core evidence comprised studies that directly applied computational methods to infant brain function. Contextual studies were retained when they established key developmental findings, biological validation targets, clinical relevance, or methodological foundations needed to interpret deep-learning applications. Evidence was synthesized according to the scientific questions addressed, the functional representations and modeling strategies used, and their developmental, biological, and translational implications. Finally, all candidate records underwent manual full-text screening to confirm eligibility, establishing a focused set of primary empirical studies for in-depth synthesis.

Whereas prior reviews have summarized infant resting-state fMRI development (Zhang et al., 2019) or deep learning across fetal, infant, and toddler neuroimaging more broadly (Chin et al., 2026), the present study takes a focused functional-neuroimaging perspective. Its central question is how **deep learning advances the neuroimaging-based analysis of infant brain function**: from how fMRI signals are represented, to how functional organization and developmental trajectories are learned, and finally to how these models can be validated, interpreted, and translated under infant-specific constraints.

## 3. Infant fMRI Data and Representations

### 3.1. Infant-specific acquisition, preprocessing, and data-quality challenges

Unlike adult fMRI, which typically relies on awake, compliant participants who can tolerate relatively long scan sessions, infant fMRI is heavily constrained by developmental status, behavioral state, safety considerations, and limited tolerance for the scanning environment. Neonates and young infants are therefore commonly scanned without sedation during natural sleep, using age-tailored procedures such as feeding before the scan, swaddling or immobilization, acoustic attenuation, thermal support, and continuous visual or physiological monitoring (Dean et al., 2014, Copeland et al., 2021). Resting-state fMRI has consequently become the dominant modality for studying the infant functional connectome because it does not require sustained attention or task compliance and is well suited to repeated longitudinal acquisition (Smyser et al., 2011, Smyser et al., 2010). In contrast, task-based fMRI is generally restricted to short, passive, and developmentally appropriate paradigms (Ellis et al., 2020, Agyeman et al., 2023).

Preprocessing must likewise be developmentally tailored. Conventional adult preprocessing pipelines assume relatively mature anatomy, stable gray-white matter contrast, and reliable registration to standardized adult templates. These assumptions do not hold in infancy, when tissue contrast changes rapidly, partial-volume effects are pronounced, brain size and cortical folding evolve substantially, and anatomical and functional boundaries remain developmentally dynamic. These features necessitate specialized infant-specific templates, tissue segmentation tools, surface-reconstruction pipelines, and functional preprocessing frameworks. For example, the dHCP neonatal fMRI pipeline was designed to address low and variable tissue contrast, substantial motion, slice-to-volume displacement, dynamic susceptibility distortion, ICA-based denoising, and automated quality control (Fitzgibbon et al., 2020). Infant-specific anatomical frameworks, such as iBEAT V2.0, further support developmentally appropriate tissue segmentation, anatomical alignment, and quality assessment before functional signals and representations are derived (Wang et al., 2023b, Shen et al., 2023).

While these advances substantially refine data preparation, they cannot fully resolve the fundamental measurement challenges intrinsic to infant fMRI. Usable acquisitions are often short or fragmented, interrupted by head motion, state transitions between sleep and wakefulness, feeding, or distress. Crucially, motion is not merely random measurement noise, as it may systematically covary with age, behavioral state, developmental status, or clinical characteristics (Badke D'Andrea et al., 2022, Kim et al., 2023a). Similarly, sleep, wake, and, in occasional clinical contexts, sedated acquisitions may reflect distinct neural states, whereas spatial normalization is complicated by rapid changes in brain size, tissue contrast, cortical folding, and functional topography (Mongerson et al., 2017, Hu et al., 2021). As a result, infant fMRI presents computational models with inputs in which neural signal, developmental stage, acquisition state, and artifacts are tightly coupled. Infant fMRI should therefore not be treated as a smaller or noisier version of adult fMRI, but as a distinct developmental measurement problem that requires models to account explicitly for data quality, state dependence, anatomical change, and uncertainty.

## 3.2 Functional data representations

Before applying any computational model, raw fMRI signals must be transformed into an analytic input representation. This step imposes a strong inductive bias regarding what counts as a functional unit, which information is preserved, and how brain function is compared across individuals, ages, and acquisitions. The representation framework can be organized along three interconnected axes: spatial granularity ranging from region-of-interest (ROI)-wise averaged to vertex- or voxel-wise dense inputs, temporal structure spanning static summaries to dynamic time series, and modal integration extending from single-modality functional inputs to structure-function or phenotype-informed representations.

The dominant input representation in infant deep-learning studies is the ROI-wise static functional connectivity (FC) matrix. Constructed by averaging blood oxygenation level dependent (BOLD) signals within predefined regions, static FC collapses a scan into a time-invariant connectome summary. Pairwise FC matrices remain widely used because they are compact, interpretable, and compatible with diverse deep architectures (Hu et al., 2020a, Xia et al., 2026b). Related static representations can reframe FC into alternative mathematical formats, such as functional principle gradients capturing continuous axes of macroscale organization(Vos de Wael et al., 2020), local gradient map emphasize local functional transition boundaries (Yuan et al., 2024), and graph-based formulations treating regions as nodes and connections as weighted edges (Li et al., 2022, Pistos et al., 2025). Despite their popularity, static ROI-level representations share a fundamental limitation: they collapse moment-to-moment network reconfigurations and risk blending fluctuating brain states into a simple temporal average.

Dynamic ROI-wise representations address this temporal limitation by preserving time-varying information, either as regional BOLD time series or as dynamic FC (dFC). These dynamic formats are conceptually attractive for isolating maturational shifts, arousal fluctuations, motion artifacts, and state variations. However, they are not yet mainstream in infant fMRI deep learning, largely because reliably estimating time-varying connectivity from short, motion-contaminated infant scans with low signal-to-noise ratios remains exceptionally challenging. Adult dFC studies demonstrate both the richness of time-varying connectivity and its sensitivity to noise, sliding-window choices, arousal state, and preprocessing pipelines (Hutchison et al., 2013, Allen et al., 2014). These vulnerabilities are amplified in infant cohorts, where empirical dynamic fMRI studies remain comparatively scarce.

When predefined regional boundaries in atlas-based approaches are insufficient or biologically restrictive, dense spatial representations preserve finer topographical detail. Vertex-wise surface representations map functional signals directly onto cortical surface meshes, maintaining local adjacency along the folded cortex and enabling fine-grained parcel-free analyses. Voxel-wise volumetric representations, in contrast, retain native three-dimensional spatial geometry for image-level representation learning. Despite demonstrated success of 3D convolutional neural networks (CNNs) in fMRI studies of subjects with older ages (Vu et al., 2020, Thomas et al., 2020), applying these models to infant cohorts remains challenged by registration sensitivity, rapid age-dependent tissue contrast changes, and marked anatomical variation in early cortical folding.

Matrix-factorized and component representations (ICA, non-negative matrix factorization (NMF), Laplacian embeddings, diffusion maps) provide another route between raw signals and deep models by transforming high-dimensional fMRI into latent components, supporting network discovery and deep-learning denoising (Doria et al., 2010, Kam et al., 2019, Heo et al., 2022, Mohapatra et al., 2025). Functional gradients, although not deep-learning methods by themselves, offer effective representations for deep models, capturing gradual functional transitions without hard parcel boundaries (Margulies et al., 2016, Yuan et al., 2024).

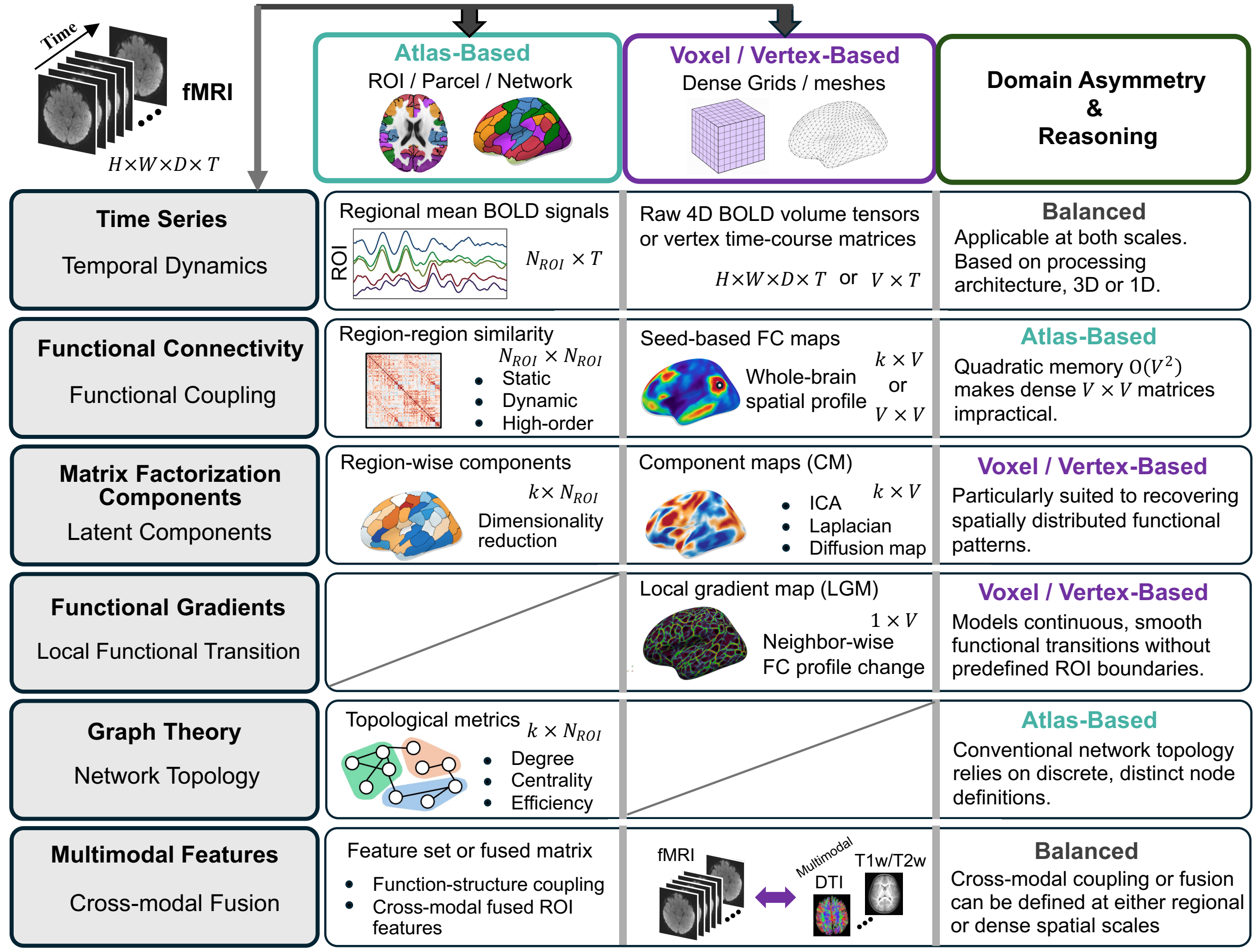

**Figure 2. Functional representations derived from infant fMRI data.** Infant fMRI representations are organized along two dimensions: spatial scale, separating **atlas-based** ROI/parcel/network representations from dense **voxel/vertex-based** volume or surface representations, and data format, including time series, functional connectivity, matrix-factorization components, functional gradients, graph-theory features, and multimodal features. The raw fMRI signal is denoted as $H \times W \times D \times T$, where $H$, $W$, and $D$ are spatial dimensions and $T$ is frame number (time points) of the fMRI scan. $N_{ROI}$ denotes the number of regions of interest, $V$ denotes the number of voxels or vertices, and $k$ represents the number of seeds, the dimension of gradients, latent components or graph features. The rightmost column indicates whether each representation is primarily atlas-based, voxel/vertex-based, or applicable at both scales (Balanced).

Multimodal representations are one of the highest-potential directions for infant fMRI deep learning because early functional organization is closely coupled with brain structure, myelination, and clinical risks. The central question is how different modalities are fused to provide complementary constraints on noisy, developmentally immature fMRI signals. One strategy is latent disentanglement, in which deep models separate modality-shared developmental information from modality-specific variation (Hu et al., 2020b, Cheng et al., 2023). A second approach is prior-guided fusion, where structural information directly shapes functional learning (Mao et al., 2024). A third strategy is generative cross-modal prediction, in which structural morphometry and age are used to reconstruct missing longitudinal FC (Zhu et al., 2025). Together, these approaches help transition infant fMRI modeling from isolated functional prediction toward biologically informed or constrained representation learning.

Ultimately, upstream representation choices shape what a deep-learning architecture can discover. An FC matrix constructed from an adult template, an age-specific infant atlas, an individualized parcellation, a continuous gradient, or a dense surface mesh encodes fundamentally different biological assumptions. **Figure 2** provides a conceptual map of how functional representations differ across atlas-based and voxel/vertex-based formulations, whereas **Table 1** summarizes the information each representation preserves, its major limitations, its relevance for deep-learning models, and representative studies. This framing emphasizes that model performance and biological validity depend not only on the neural network architecture, but also on how infant fMRI is represented before learning begins.

**Table 1. Functional data representations for deep learning models.**

| Representation family | Information preserved | Main limitation / caution | Matched deep-learning models | Representative studies / Context |
|---|---|---|---|---|
| **Time series / Functional dynamics** | Temporal order, BOLD fluctuations, dynamic state transitions | Vulnerable to motion, physiological noise and age-dependent BOLD-response delays; memory intensive for long sequence modeling. | 1D CNN, RNN / LSTM / GRU, TCN, Temporal Transformer | (Wu et al., 2025, Heo et al., 2022, Heo et al., 2025) / (Cheng et al., 2025b, Li et al., 2020) |
| **Functional connectivity** | Region-to-region coupling, network organization, individual FC fingerprints | Collapses temporal dynamics; highly sensitive to atlas choice and node definition | MLP / fully connected deep neural network (DNN); AE / VAE; 2D CNN or U-Net; Transformer / masked AE | (Xia et al., 2026b, He et al., 2018, Hu et al., 2020a, Mao et al., 2024, Xia et al., 2026a, Hu et al., 2021) |
| **Matrix factorization components** | Distributed latent networks, signal/noise components, low-dimensional functional modes | Inconsistent component identity/order across cohorts, preprocessing pipelines, and model choices. | CNN plus RNN/LSTM/GRU hybrids; attention-based spatiotemporal networks; Transformer plus factorization hybrids | (Kam et al., 2019, Heo et al., 2022, Heo et al., 2025, Mohapatra et al., 2025) / (Kim et al., 2023a) |
| **Functional gradients** | Continuous functional axes and local changes in FC profiles | Susceptible to manifold alignment instability, registration errors, and neighborhood definition; infant deep learning use is still sparse. | Spherical U-Net / surface CNN; contrastive surface networks; gradient-informed representation learning | (Cheng et al., 2025a)/ (Yuan et al., 2024, Hu et al., 2024) |
| **Graph theory / network topology** | Node-edge topology, integration/segregation metrics, network hubs, modularity | Graph findings frequently reflect artificial matrix thresholding, edge definition rules, or atlas parcellation rather than biology. | GNN / GCN / GAT; spatiotemporal graph Transformer; federated GNN | (Li et al., 2022, Pistos et al., 2025, Pistos et al., 2023, Bessadok et al., 2021) |
| **Multimodal features** | Structure-function coupling, myelination, morphology, clinical / behavioral context | Prone to missing modalities, cross-site harmonization bias, cohort imbalance, and developmental age-confounding. | Multimodal/adversarial autoencoder; dual-branch CNN; SC-guided Transformer; Cross-modal attentive diffusion model | (Hu et al., 2020b, Cheng et al., 2023, Mao et al., 2024, Zhu et al., 2025, Luo et al., 2026) / (Huang et al., 2023) |

Note: In the "Representative studies / Context" column, references before the slash (or single citations with no slash) refer to infant fMRI deep learning or related computational studies, and those after the slash indicate broader context, such as non-infant, non-deep-learning, or non-fMRI studies. This format applies to Tables 1 through 4. Abbreviations: D, dimensional; CNN, convolutional neural network; RNN, recurrent neural network; LSTM, long short-term memory; GRU, gated recurrent unit; TCN, temporal convolutional network; MLP, multilayer perceptron; DNN, deep neural network; AE, autoencoder; VAE, variational autoencoder; GNN, graph neural network; GCN, graph convolutional network; GAT, graph attention network; SC, structural connectivity.

### 3.3 Deep Representation Learning of Infant Functional Data

Once infant fMRI has been acquired, preprocessed, and formatted into analytic representations, deep learning can further project these inputs into latent embeddings or low-dimensional manifolds. Rather than treating functional features as static inputs, deep models learn task-adaptive representations that may emphasize developmental maturation, individual-specific organization, network topology, or multimodal structure. Deep representation learning thus bridges raw infant fMRI representations and downstream clinical or cognitive objectives.

As summarized in **Figure 3** and **Table 2**, fMRI deep-learning architectures can be categorized by the input representations and underlying inductive biases. Vector-based models, e.g., multilayer perceptrons (MLPs) or fully connected networks, treat functional features as unconstrained vectors, allowing arbitrary feature interactions without assuming spatial or temporal order. Spatial models, e.g., convolutional neural networks (CNN) and U-Nets, assume local spatial neighborhoods contain meaningful structure that can be hierarchically combined, whereas spherical U-Nets extend these principles to cortical surface geometry. Graph-based models, e.g., graph neural networks (GNN) and graph convolutional network (GCN), formalized brain organization as network topology, propagating information along defined anatomical or functional edges. Attention-based model, e.g., transformers, assume that long-range dependencies can be learned through attention across regions, patches, visits, or modalities. Temporal models, e.g., recurrent neural network (RNN), Long short-term memory network (LSTM), and gated recurrent unit (GRU), enforce sequential ordering to capture continuous BOLD dynamics or longitudinal developmental trajectories.

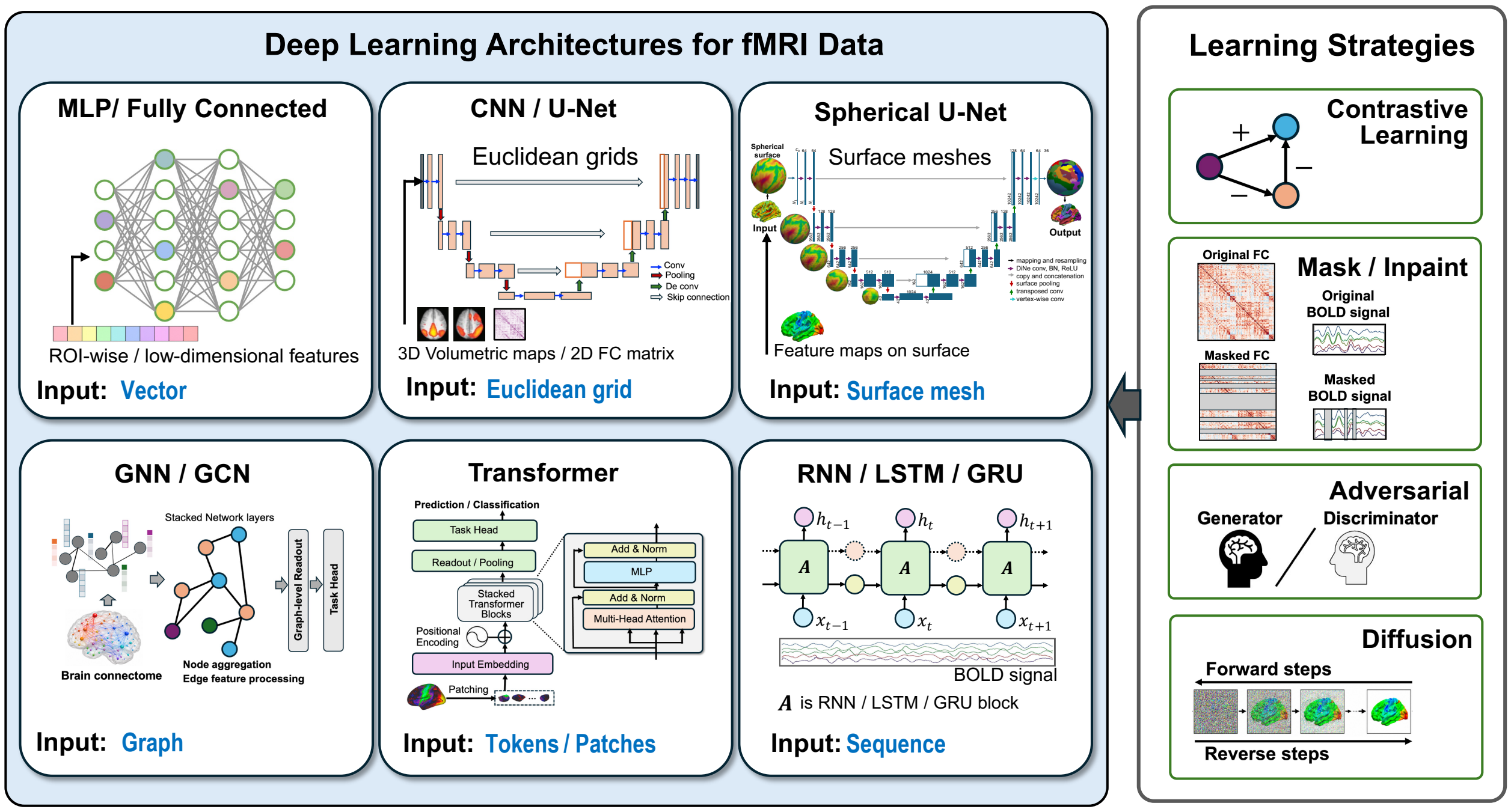

**Figure 3. Deep-learning architectures and training strategies for infant fMRI data.**

Learning strategies further shape what an architecture learns from infant functional data. While supervised learning excels when reliable behavioral or clinical targets are available, it remains susceptible to overfitting in small, noisy infant cohorts. Self-supervised learning offers a powerful alternative by leveraging unlabeled scans. Contrastive learning encourages representations to remain consistent across biologically meaningful augmentations, visits, subjects, or modalities, whereas masked modeling trains the model to recover occluded parts of the input, such as FC edges, BOLD segments, image patches, surface patches, or even masked latent representations. Adversarial learning enforces domain-invariant representations through a minimax game, stripping away site and acquisition confounds to improve model robustness and cross-site generalization. Diffusion-based strategies offer a way to synthesize high-fidelity functional states and forecasting developmental trajectories through iterative reverse denoising.

For infant fMRI, architectural selection should be guided by scientific objectives and input representations rather than model capacity. Because acquisition confounds and genuine biological signals are often tightly intertwined in early life, over-parameterized models are highly vulnerable to shortcut learning, that is, achieving spuriously elevated performance by exploiting cohort-specific confound structures rather than true neurodevelopmental information. Consequently, robust infant fMRI modeling requires not only an objective-aligned data representation and model architecture, but also age-aware sampling, sensitivity analyses, external validation, uncertainty quantification, and biological interpretability.

**Table 2. Deep-learning architectures for infant fMRI.**

| Architecture | Main advantage | Infant-specific caution | Representative studies / Context |
|---|---|---|---|
| **MLP / fully connected DNN** | Efficient for compact prediction from ROI-wise or low-dimensional functional features. | Limited ability to preserve spatial topology or network structure; prone to shortcut learning in small cohorts. | (He et al., 2018, Ali et al., 2022) |
| **CNN / U-Net** | Captures local and hierarchical spatial patterns in image-like fMRI inputs. | Local neighborhoods may be artificial in reordered FC matrices; volumetric infant fMRI remains limited by registration and dimensionality. | (Kam et al., 2019, Heo et al., 2022, Heo et al., 2025) / (Vu et al., 2020); |
| **Spherical U-Net / surface CNN** | Preserves cortical-surface topology for fine-grained functional mapping. | Sensitive to surface registration, mesh resolution, and age-related changes in cortical folding. | (Cheng et al., 2025a) / (Hu et al., 2024, Zhao et al., 2021) |
| **GNN / GCN** | Models connectome topology through node-edge message passing. | Findings depend strongly on atlas choice, node definition, edge construction, and graph sparsity. | (Li et al., 2022, Pistos et al., 2025, Zhang et al., 2022) / (Speckert et al., 2025) |
| **Transformer** | Captures long-range dependencies across regions, time windows, dFC states, visits, or modalities. | Data-hungry and vulnerable to shortcut learning from site, age, missingness, or scan quality. | (Mao et al., 2024, Fang et al., 2023, Wu et al., 2025, Cheng et al., 2025b) |
| **RNN / LSTM / GRU** | Models temporal ordering in BOLD sequences, time windows, or longitudinal visits. | Short scan duration, motion bursts, arousal changes, and sleep-state variation can mimic neural dynamics. | / (Cheng et al., 2025b, Dvornek et al., 2017) |

## 4. Deep Learning for Functional Mapping

Functional mapping aims to delineate the spatial and topological organization of brain function, including functional networks, gradients, connectome fingerprints, and individualized areal boundaries. In infancy, this objective is especially important because functional architecture reorganizes rapidly across development, rendering adult-derived group templates inadequate for capturing age-specific or subject-specific topologies. While conventional approaches, including ICA, atlas-based parcellation, and template matching, have established the broad developmental landscape of infant functional organization, they typically depend on fixed features, group averaging, or linear assumptions. Deep representation learning offers a paradigm shift by extracting nonlinear, high-dimensional and topology-aware features from noisy infant fMRI, enabling age-conditioned and individualized modeling of early functional architecture.

## 4.1. Learning population-level functional organization

Rather than relying on group-average FC maps or linear similarity metrics, deep models project high-dimensional infant fMRI data onto continuous, low-dimensional manifolds that bring functionally congruent voxels, vertices, or parcels together while segregating distinct functional units. Functional systems can subsequently be delineated from these learned manifolds via clustering, latent component extraction, or spatial boundary detection. Moreover, explicitly conditioning deep models on key covariates related to network formation, such as age and scan state, may help distinguish intrinsic functional architecture from maturational trajectories and state-dependent fluctuations.

Empirical applications of deep learning to population-level infant functional organization remains limited but emerging. Kim et al. demonstrated that variational autoencoder (VAE)-derived latent variables from fetal and neonatal rs-fMRI capture shared developmental structure and decode distributed functional network patterns (Kim et al., 2023b). Mohapatra et al. extracted Transformer-derived features to represent distributed functional relationships and leveraged matrix factorization to obtain group-level neonatal network maps (Mohapatra et al., 2025). These studies highlight that the core utility of deep learning in population mapping lies in learning nonlinear functional embeddings from which emerging canonical systems and population-level developmental patterns can be derived.

## 4.2. Learning individualized functional organization

Individualized functional mapping addresses a complementary challenge to identify functional signatures that remain stable and distinctive within an individual infant. This is essential for early developmental neuroscience because individual functional organization may provide the basis for later cognitive, behavioral, or clinical differences. However, individual mapping remains difficult due to rapid functional development, short scan duration, and motion artifacts.

Deep learning addresses these constraints by learning latent spaces that explicitly separate subject-specific signatures from developmental variation. For instance, a disentangled intensive triplet autoencoder has been used to evaluate connectome fingerprinting in infancy, isolating identity-specific features from age-dependent characteristics to improve subject separability (Hu et al., 2020a). Kim et al. extended this idea in newborns by using a VAE to learn latent representations of neonatal functional connectomes, showing that nonlinear latent variables can distinguish age-related developmental features from individual-uniqueness signatures more effectively than linear models (Kim et al., 2024).

Beyond atlas-based FC matrices, deep learning also enables spatial localization of subject-specific boundaries. Tu et al. supports this direction by using deep-learning-based embeddings of functional connectivity profiles for precision functional mapping (Tu et al., 2025). Surface-based deep models, such as spherical U-Net and contrastive learning frameworks, preserve mesh topology and local spatial adjacency on the cortical sheet, facilitating fine-grained functional boundary estimation as well. While initially validated in structural morphometry (Zhao et al., 2019, Zhao et al., 2021) or adult fMRI cohorts (Hu et al., 2024), these surface-based deep boundary mapping frameworks extend naturally to early life.

While these studies highlight the potential of deep learning for individualized functional-organization analysis, the central caution is that the learned individual signatures must reflect stable functional architecture rather than acquisition artifacts, scan-state effects, site-related bias, or age-related confounding. Individualized deep mapping will therefore be convincing only when it demonstrates test-retest reliability, cross-session stability, age-aware interpretation, and biological plausibility.

## 5. Modeling Developmental Dynamics and Individual Trajectories

Traditionally, infant functional brain development has been characterized using classical statistical frameworks, including regression models, mixed-effects models, generalized additive models, and growth-curve approaches. These methods have illuminated key age-related changes in functional networks, topological segregation and integration, hub organization, and regional developmental gradients (Smyser et al., 2010, Cao et al., 2017, Wen et al., 2019, Jiang et al., 2023, Li et al., 2024, Liu et al., 2023). While their primary strength lies in high interpretability and rigorous statistical inference, these classical approaches typically rely on predefined imaging features and population-average trajectory models. Consequently, they are inherently limited in their capacity to resolve high-dimensional, nonlinear, and subject-specific dynamics of functional maturation. Deep learning marks a fundamental methodological shift, moving beyond group-average descriptive trajectories toward estimating an individual infant's maturational state and forecasting personalized neurodevelopmental trajectories over time.

### 5.1. Brain functional maturation estimation

Brain-age prediction serves as the main computational framework for estimating individualized functional maturation. By projecting high-dimensional neuroimaging features to chronological, gestational, or postmenstrual age, deep learning models position an individual brain along a data-driven maturational axis. The resulting brain age gap (BAG), the discrepancy between predicted and observed age, provides a quantitative index of accelerated or delayed functional development. This paradigm is uniquely compelling in infancy, where functional connectivity reorganizes rapidly across weeks and months, offering the potential to detect subtle neurodevelopmental deviations before overt clinical or behavioral phenotypes manifest.

Classical machine-learning studies first demonstrated that infant functional connectomes encode rich developmental signatures (Smyser et al., 2016, Sun et al., 2024). Deep-learning architectures extend this foundation by learning complex, nonlinear, and distributed interactions directly from high-dimensional connectome representations. For example, Li et al. used a GCN framework for infant age prediction from FC, representing brain regions as nodes and functional relationships as graph edges (Li et al., 2022). Multimodal approaches provide another route when complementary imaging information is available. Hu et al. proposed a disentangled multimodal adversarial autoencoder for infant age prediction in the presence of incomplete structural MRI and fMRI data, addressing a frequent challenge in longitudinal infant imaging in which modalities may be missing or differ substantially in quality (Hu et al., 2020b).

While brain-age modeling summarizes complex functional topographies into a single maturational metric, predicted age should not be naively equated with true biological maturity. High-capacity models can easily exploit non-biological correlates of age, such as head motion, scan-state shifts, and scanner protocol variations. Brain-age deviation therefore requires careful interpretation, external validation, uncertainty quantification, and demonstration that the learned representation reflects biologically meaningful developmental variation rather than acquisition-related confounds. Looking forward, deep normative modeling represents a critical frontier. Rather than collapsing multidimensional functional organization into a single scalar age, generative normative frameworks can learn age- and covariate-conditioned distributions of normative functional development, enabling precise, multi-index mapping of individualized deviations with calibrated statistical confidence.

## 5.2. Functional-connectome forecasting and individual trajectory prediction

Whereas brain-age models estimate an infant's current brain maturational status, longitudinal forecasting models aim to predict how functional organization evolves over time. Their objectives include predicted future functional connectomes or imputed missing longitudinal observations. This represents a significantly more complex task, requiring models to minimize reconstruction error while simultaneously preserving network topology, developmental properties, and individual-specific functional organization.

One primary direction is FC-only trajectory forecasting, which uses single or multiple FC observations to predict missing or future FC states. Yu et al. introduced a conditional intensive triplet network for longitudinal infant FC prediction, explicitly enforcing the preservation of subject-specific connectivity characteristics (Yu et al., 2022b). Leveraging self-supervised representations, Xia et al. developed a triplet longitudinal masked autoencoder to reconstruct individualized functional connectome trajectories from incomplete longitudinal scans, combining connectivity-specific representation learning with constraints on developmental continuity and individual identity (Xia et al., 2026b).

A complementary direction leverages multimodal or cross-modal information to guide functional trajectory prediction. Bessadok et al. proposed a few-shot graph multi-trajectory evolution network that predicts multiple infant connectivity trajectories from a baseline observation (Bessadok et al., 2021), enabling functional and morphological connectome development to be modeled jointly. Extending this approach to multi-site data, Pistos et al. introduced federated multi-trajectory GNN to aggregate developmental trajectories across institutions while maintaining data privacy (Pistos et al., 2023). Moreover, Zhu et al. similarly used structural morphometry and target age to guide longitudinal FC generation (Zhu et al., 2025), whereas Luo et al. explored cross-modal synthesis to generate fMRI-derived FC from EEG when fMRI is difficult to acquire (Luo et al., 2026).

The central challenge in connectome forecasting is moving beyond element-wise loss metrics (such as mean squared error) toward topological and biological validation. Low reconstruction loss does not guarantee that a predicted connectome preserves global network topology, spatial developmental gradients, subject-specific functional fingerprints, or clinically meaningful trajectories. Comprehensive evaluation of forecasting models must therefore incorporate topological graph metrics, individualized functional fingerprint, and longitudinal stability checks.

## 6. Translational Applications, Model Evaluation, Interpretability, and Robust Learning

Deep learning holds immense translational potential in infant fMRI, offering the capacity to link early functional brain organization to long-term cognitive outcomes, clinical risks, environmental exposures, and individualized neurodevelopmental trajectories. However, for a deep-learning framework to achieve genuine clinical utility in infant neuroimaging, it must satisfy stringent methodological criteria: generalize reliably across heterogeneous acquisition sites, yield calibrated uncertainty estimates, provide biologically grounded explanations, and remain resilient against intrinsic infant-imaging artifacts and sampling constraints. In this section, we examine translational applications, model evaluation, biological interpretability, and robust representation learning not as isolated methodological milestones, but as deeply interdependent prerequisites. Ultimately, a deep learning model is clinically actionable in early life only when its predictions are simultaneously accurate, robust, mechanistically transparent, and developmentally plausible.

### 6.1 Translational applications in clinical and developmental contexts

Translational applications ask whether early functional organization can help predict or characterize outcomes that matter for development and clinical care. In infant fMRI, deep learning is particularly attractive because early pathological and maturational signals are typically subtle, nonlinear, and distributed across complex connections rather than isolated brain regions. While functional maturity estimated as brain-age is one important translational target (discussed in Section 5), this subsection focuses mainly on cognitive/behavioral prediction and prematurity/birth-related risk (**Table 3**).

**Table 3. Translational targets of deep learning in infant functional neuroimaging**

| Application domain | Prediction / modeling target | Key translational caution | Representative studies / Context |
|---|---|---|---|
| **Functional maturity** | **Regression**: postmenstrual or chronological age, brain-age gap, maturational state. **Classification**: delayed versus typical maturation. | Brain age becomes translational only when deviation is calibrated against clinical context and developmental outcomes. | (Kim et al., 2023b, Li et al., 2022, Hu et al., 2020b) / (Smyser et al., 2016, Sun et al., 2024, Brown et al., 2017) |
| **Cognitive and developmental outcome prediction** | **Regression**: cognitive, language, motor, or social-emotional scores. **Classification**: neurodevelopmental deficit or high-risk outcome. | Small long-term outcome cohorts make shortcut learning easy; models must outperform clinical baselines. | (He et al., 2018, Ali et al., 2022, He et al., 2021, Cheng et al., 2023, Wu et al., 2025) |
| **Prematurity and birth-related risk characterization** | **Regression**: gestational age, risk severity, or altered-connectivity burden. **Classification**: preterm versus term or risk-group status. | Models may learn medical-care, scanner, or injury signatures instead of prematurity-specific functional organization. | (Zhang et al., 2022, Mao et al., 2024) / (Ball et al., 2016, Li et al., 2025) |
| **Exposure, environment, and caregiving context** | **Regression**: exposure level, caregiving quality, socioeconomic measures, or behavioral scores. **Classification**: exposed versus non-exposed groups. | Contextual variables are deeply confounded; deep learning evidence remains mostly potential rather than mature. | / (Graham et al., 2015, Ramphal et al., 2020, Phillips et al., 2021, Merhar et al., 2021, Banihashemi et al., 2023) |
| **Clinical-risk conditions** | **Regression**: symptom severity or continuous clinical scores. **Classification**: diagnosis, risk group, or treatment/prognosis category. | Condition-specific datasets are small, so calibrated risk estimates and external replication matter more than within-cohort accuracy. | / (Rogers et al., 2017, Yu et al., 2022a, Alotaibi et al., 2022, Yin et al., 2024) |

The most developed deep learning application is prediction of cognitive or neurodevelopmental outcomes, particularly in very-preterm infants. MLP-based and semi-supervised models have been used to predict cognitive deficits from neonatal or early-life FC, showing how deep learning can integrate subtle, distributed features even when labels are scarce and outcome phenotypes are heterogeneous (He et al., 2018, Ali et al., 2022). Multimodal deep-learning models further enhance predictive performance by combining fMRI-related features with structural MRI, diffusion tractography, cortical morphometry, or clinical metrics, reflecting the incremental contribution of complementary modalities to fMRI on cognition prediction (He et al., 2021, Cheng et al., 2023, Wu et al., 2025). In this setting, the primary translational value of deep learning lies in synthesizing multi-system, multimodal representations rather than isolating single predictive connections.

Prematurity and birth-related risk modeling aims at using deep learning to characterize the functional-network alterations associated with early biological risk. While conventional machine-learning-based fMRI studies have shown that preterm birth, neonatal encephalopathy, and perinatal risk are associated with altered functional connectivity (Ball et al., 2016, Li et al., 2025, Alotaibi et al., 2022), deep learning offers a way to model these alterations as distributed network patterns. Multimodal GNN has been used to differentiate preterm and term infant functional connectomes and characterize altered connectivity patterns, whereas structural-connectivity-guided Transformers utilize anatomical connectivity to constrain the identification of preterm-related functional abnormalities associated with prematurity (Zhang et al., 2022, Mao et al., 2024).

These studies illustrate a direction toward learning multivariate biomarkers of altered early functional organization.

Existing infant fMRI studies have demonstrated that functional organization is sensitive to prenatal exposures, socioeconomic context, caregiving, and clinical risks, yet the literature in these domains is still dominated by conventional association, mediation, and machine learning analysis. Deep learning may become valuable when it can integrate high-dimensional FC with longitudinal behavior, environmental measures, clinical variables and multimodal neuroimaging, but direct infant fMRI deep learning evidence remains limited. **Table 3** therefore presents these translational targets while distinguishing them from domains with stronger deep-learning evidence. Overall, translational deep learning in infant fMRI is most compelling when it uses distributed functional organization to predict or characterize complex outcomes that cannot be captured by isolated FC edges or conventional covariates alone. At present, the strongest evidence lies in functional maturity estimation, cognitive outcome prediction and prematurity-related risk characterization, whereas environmental and disorder-specific applications represent high-potential but less established.

### 6.2 Model and Biological Validation

Rigorous validation determines whether deep learning models applied to infant fMRI captured genuine neurodevelopmental signals or merely reflect spurious dataset artifacts. Conventional evaluation often reports accuracy, correlation, area under the receiver operating characteristic curve (ROC-AUC), or mean prediction error, but these metrics are usually insufficient for high-dimensional neurodevelopmental representations. In infant cohorts, model performance can be inflated by longitudinal repeated scans, small sample sizes, site effects, missing visits, and age-correlated motion. Consequently, a model may perform by exploiting shortcuts such as acquisition artifacts, preprocessing bias, or cohort-specific demographic biases.

Validation frameworks must therefore align with the scientific claim being made. For classification tasks, such as clinical diagnosis or risk-group stratification, classic accuracy and ROC-AUC should be complemented by metrics robust to class imbalance, such as precision-recall AUC (PR-AUC), balanced accuracy, or macro-F1. For continuous regression tasks, such as brain-age estimation or cognitive-score prediction, global summary statistics should be supplemented by age-stratified errors, residual analyses, and controlled sensitivity testing for motion, gestational age, sex, site, and other relevant covariates. For connectome forecasting, evaluation should extend beyond element-wise reconstruction error to test whether predicted connectomes preserve macroscale topology, individual fingerprints, and neurobiologically plausible developmental trajectories. Finally, for self-supervised representation-learning models, latent spaces should be evaluated for topological stability, probabilistic calibration, and sensitivity to confounds rather than only downstream prediction accuracy.

External validation is especially important in infant fMRI. Held-out subjects are necessary but not sufficient when testing within the same scanner, acquisition protocol, or preprocessing pipeline.

Robust evidence demands out-of-distribution evaluation across held-out sites, independent prospective cohorts, age-stratified test splits, and multi-center or federated evaluation designs that reduce the risk of site-specific overfitting (Pistos et al., 2023, Pistos et al., 2025). Beyond empirical generalizability, biological validation evaluates whether learned features correspond to plausible neurodevelopmental mechanisms. Model-derived latent embeddings, attention patterns, saliency maps, or important edges should be compared with independent multimodal evidence such as infant-specific functional networks (Wang et al., 2023a), structural and diffusion MRI (Girault et al., 2019, Hong et al., 2023), myelination-sensitive measures (Huang et al., 2023), and prospective behavioral outcomes.

A practical validation hierarchy can therefore be organized into three levels: methodological integrity, external generalization, and biological convergence. The first requires subject-level splitting, leak-free preprocessing, appropriate baselines, and uncertainty reporting. The second requires held-out-site or prospective replication, age-stratified performance profiling, and comprehensive confound sensitivity analyses. The third requires evidence that learned representations align with multimodal biological evidence. Under this hierarchical framework, infant fMRI deep learning advances from showing raw predictive capacity to proving that it captures developmentally meaningful brain function.

### 6.3 Interpretability of deep infant functional models

Conventional infant fMRI interpretation relies heavily on predefined brain parcellations, graph-theoretic topologies, or direct statistical associations with age and behavior, where input features inherently retain explicit biological meaning. In contrast, deep architectures excel at projecting high-dimensional inputs into abstract, non-linear latent spaces, yielding complex spatiotemporal representations that are not inherently self-explanatory. In infant neuroimaging, interpretability is therefore not merely a tool for explaining predictions, but a critical diagnostic mechanism to test whether a model has captured plausible neurodevelopmental biology.

Several complementary approaches can support this goal. Feature-attribution and perturbation methods can identify FC edges, regions, temporal windows, or anatomical locations that most influence a prediction. Attention profiles in Transformers and node- or edge-level weights in graph models may highlight candidate networks or developmental features, though attention weights should be treated as dynamic information routing rather than standalone explanations. Latent-space visualization and disentanglement analyses can test whether learned dimensions isolate biological variables from non-biological confounders. While both inherently interpretable architectures and post-hoc explanation tools offer valuable insights into model behavior, all derived contributive feature maps should be treated mainly as hypothesis-generating rather than mechanistic proof (Bass et al., 2023, Wu et al., 2024).

Useful interpretation should be stable, developmentally coherent, and biologically convergent. Feature attributions must remain robust against resampling, perturbation, and sanity checks. Feature importance should evolve in age-appropriate ways, reflecting known non-linear trajectories of brain maturation. Inferred networks and brain regions should align with independent cross-modal benchmarks. Representational-alignment approaches offer another useful framework by comparing infant neural activity patterns with the features learned by deep neural networks (O'Doherty et al., 2026), but such similarity should be interpreted cautiously because computational alignment does not necessarily imply shared biological mechanism.

### 6.4 Robust learning under limited and heterogeneous infant fMRI data

Robust learning represents a critical domain where deep learning offers distinct methodological advantages for infant fMRI. Conventional preprocessing and artifact correction strategies, such as frame censoring, nuisance regression, ICA-based denoising, ComBat-style harmonization, and covariate adjustment, remain essential. However, these traditional tools often struggle when technical variability is nonlinear, developmentally structured, and intrinsically coupled with underlying biology. In infant cohorts, acquisition artifacts and sampling constraints frequently covary with gestational age, postmenstrual age, scan state, and clinical status. Linear corrections therefore risk leaving hidden interactions intact or inadvertently stripping away meaningful developmental variance. **Table 4** summarizes the major variability targets, modeling risks, validation priorities, and representative studies.

**Table 4. Infant-specific fMRI data challenges and robust learning**

| Data challenge | Main risks | Validation priority | Representative studies / Context |
|---|---|---|---|
| **Motion artifacts and denoising** | Residual head motion or physiological noise mimics age, diagnosis, or identity; aggressive denoising strips developmental signal. | Report motion; use motion-matched and sensitivity analyses; test whether correction preserves age effects, fingerprints, and downstream associations. | (Kam et al., 2019, Heo et al., 2022, Tang et al., 2025, Xia et al., 2026a, Heo et al., 2025) |
| **Scan state variation** | Sleep, wakefulness, and arousal differences may be learned as maturation, diagnosis, or site effects. | Record and stratify scan state; validate state translation by preserving individual identity, developmental structure, and clinically relevant variation. | (Hu et al., 2021) / (Yin et al., 2025, Yates et al., 2023) |
| **Short scan duration** | Low usable frame counts degrade FC reliability, destabilize individualized mapping, and amplify dynamic FC sampling error. | Use scan-length-matched baselines; report performance by retained frames; test reliability of individual-level representations. | / (Hu et al., 2024, Luckett et al., 2023) |
| **Atlas dependence** | Adult templates or arbitrary node definitions distort infant boundaries, producing parcellation-dependent findings. | Use infant-specific templates; repeat analyses across atlases, spatial resolutions, and node definitions; test atlas-independent or surface-based alternatives. | (Xia et al., 2026a) / (Wang et al., 2023a, Wang et al., 2022, Chen et al., 2026) |
| **Site/scanner effects** | Scanner hardware, protocol, or preprocessing differences dominate latent spaces and confound age/clinical labels. | Use held-out-site testing; evaluate both nuisance reduction and preservation of trajectories, fingerprints, and outcome-relevant information. | (Xia et al., 2025, Cheng et al., 2025a, Pistos et al., 2025) / (Almuqhim and Saeed, 2025, Chan et al., 2023) |

Deep learning addresses these challenges by learning transformations that are nonlinear, representation-aware, and task-sensitive. Rather than relying on hand-crafted rules, deep noise-component detection models automatically identify complex artifact signatures (Kam et al., 2019).

Similarly, scan-state translation architectures explicitly align functional connectomes across awake and sleeping states, treating arousal dynamics as structured state transitions rather than merely as a nuisance covariate (Hu et al., 2021). Deep harmonization frameworks and neural multi-atlas models effectively attenuate scanner and motion artifacts while preserving local functional topography, longitudinal consistency, and fine-grained population organization (Xia et al., 2025, Cheng et al., 2025a, Xia et al., 2026a). Furthermore, while initially validated on adult data, contrastive surface-based paradigms designed to optimize short-scan reliability hold direct relevance for infant imaging, offering a powerful mechanism to stabilize individualized functional parcellations against the brief and interrupted acquisition windows typical of early development (Hu et al., 2024).

The central methodological challenge in deep robust learning is preventing overcorrection. Because deep architectures possess high expressive capacity, models optimized to remove motion, site, scan-state, or atlas effects risk simultaneously eliminating age-dependent, subject-specific, or clinically meaningful signal. Consequently, robustness must be evaluated against a dual-validation standard: the effective reduction of measurement artifacts coupled with the explicit preservation of developmental biology. Rigorous evaluation must verify that corrected representations retain normative age trajectories, individual connectivity fingerprints, structure-function alignment, longitudinal stability, and prospective clinical utility. Ultimately, the goal is not to impose artificial homogeneity across heterogeneous cohorts, but to disentangle technical noise from true biological variance.

## 7. Challenges and Future Directions

Deep learning has begun to transition infant functional neuroimaging from descriptive, group-averaged analyses toward individualized representation learning and predictive developmental modeling. Nevertheless, the field remains in its own infancy. Many current frameworks are proof-of-concept demonstrations built on small, single-center, demographically heterogeneous cohorts. The next phase of progress will depend less on scaling parameter counts or architectural complexity and more on building models that are developmentally grounded, externally validated, biologically interpretable, and clinically actionable.

The most immediate challenge is data scale, harmonization, and metadata transparency. While existing repositories remain constrained by abbreviated scan durations, irregular age sampling, missing longitudinal follow-ups, and pronounced site-to-site acquisition variance, large-scale, harmonized, and longitudinally dense infant fMRI datasets are much needed. But scale alone is not sufficient. Future data collection and benchmark release should systematically report motion, scan state, acquisition protocol, detailed socio-demographic context, and clinical variables in standardized ways, enabling age-stratified analyses, held-out-site testing, and out-of-distribution validation of deep learning models.

Next-generation models should move from task-isolated prediction toward richer developmental representation learning. Static FC matrices will remain useful, but future models should increasingly integrate temporal dynamics, surface topology, graph structure, multimodal imaging, and clinical context. Self-supervised and foundation-model approaches could help by pretraining on large neuroimaging datasets and then adapting to infant-specific developmental trajectories. Pretrained neuroimage models may also provide anatomical or tissue-maturation priors when structural MRI, diffusion MRI, or myelination-sensitive measures are available. However, these models must be adapted cautiously because adult or general neuroimage priors may not transfer cleanly to the rapidly changing infant brain.

A transformative opportunity lies in connecting infant fMRI deep learning with broader, unstructured clinical information. Large language models (LLMs) and vision-language architectures offer a mechanism to harmonize complex electronic health records (EHRs), including perinatal complications, maternal health history, birth weight, serial cranial ultrasounds, and standardized neurodevelopmental assessments, into structured semantic embeddings. Rather than replacing primary neuroimaging models, language-guided systems can facilitate phenotyping, cohort stratification, literature-guided model design, and clinically interpretable summarization. The strongest future systems may therefore combine infant fMRI representations with structured clinical variables, imaging-derived biomarkers, and language-derived developmental context, while preserving clear boundaries between predictive inference, diagnostic explanation, and clinical decision-making.

For deep learning to achieve real-world clinical utility in pediatric neurology, models must satisfy rigorous evidentiary standards beyond internal cross-validation accuracy. Predictive tools targeting neurodevelopmental conditions, such as autism spectrum disorder, cerebral palsy, or extreme prematurity complications, must provide calibrated uncertainty estimates and demonstrate actionable incremental value over routine clinical variables and risk scores. Concurrently, interpretability must evolve beyond post-hoc, visually ungrounded saliency maps toward circuit-level, biologically validated explainability.

Deep learning stands at a pivotal juncture in infant functional neuroimaging. By moving away from black-box, static regression or classification toward developmentally grounded representation learning and continuous trajectory forecasting, deep learning offers an unprecedented capability to decipher the complex, nonlinear, multiscale dynamics of early brain development. Overcoming current data fragility, evaluation, and interpretability bottlenecks will establish deep learning not merely as a predictive tool, but as an indispensable scientific framework for understanding early functional brain maturation and informing timely, individualized early-life interventions.

## Acknowledgements

This work was supported in part by the National Institutes of Health grants HD119560, AG075582, NS128534, EB037388 and NS135574.